# Full Band Gap and Defects States in Solid-in-Solid Three Dimensional Phononic Crystals

Ke Sun and Z Yang[1]
Department of Physics, the Hong Kong University of Science and Technology,
Clearwater Bay, Kowloon, Hong Kong
Email: phyang@ust.hk

**Abstract**
A full band gap of the longitudinal mode of elastic waves centered near 2.8 MHz with a width of ~ 1 MHz has been observed in the phononic crystals made of body centered tetragonal tungsten carbide spheres imbedded in aluminum matrix. Two defects states in the band gap due to a 7-sphere defect cluster with silicon nitride spheres have also been observed. Transmitted pressure field pattern clearly shows that at the defect state frequencies the ultrasonic waves transmitted through the doped crystal are emitted from the defect cluster.

---

[1] The author to whom any correspondence should be addressed

## 1. Introduction

Phononic crystals (PCs) are artificial structures with periodic variation of elastic properties. They have attracted much interest in the last 20 years owing to the rich physics involved and potential applications. Full three dimensional (3D) gap was predicted in PCs made of high density spheres in a low density matrix material [1]. A multiple scattering method was later developed to more accurately model the elastic wave propagation in PCs. A number of different scatterers and matrix materials, with different lattice structures and filling ratios, were investigated in search of large and full 3D gap [2 - 5]. Solid-in-solid PCs were found to exhibit larger bandgaps while relaxing the topological constraints as compared to solid-in-fluid PCs [6]. Two dimensional (2D) PCs are relatively easy to fabricate and were therefore the first systems investigated experimentally [7 - 10]. A full gap was found in many cases, and theoretical predictions were found in good agreement with the experimental results. The 3D PCs are more difficult to fabricate, and only a limited number of studies have been reported so far. A partial gap was observed in 3D PCs made of steel balls in polymethyl methacrylate (PMMA) matrix [11], in colloidal crystals [12], and in a PC made of woodpile of PMMA rods [13]. 3D PCs made of closely packed steel beads in water exhibited a full gap [14], the tunneling of ultrasound in the gap [15], and the ultrasound focusing phenomena in a higher passband [16]. Recently a full gap from 400 KHz to 700 KHz in the (100), (110), and (111) directions was observed in a PC of steel balls in epoxy matrix [17]. However, due to its high loss in the MHz range, epoxy is not an ideal matrix material for PCs in the frequency range higher than 1.5 MHz.

Defects were introduced by breaking the spatial periodicity, which can cause wave localization and act like waveguides [18]. Most of the researches on defects are on 2D PCs [14, 19 - 21]. Defects band in the gap was theoretically predicted for point defects in 3D PCs of water spheres embedded in mercury [22]. To our best knowledge, no experimental work on defects in 3D PCs has yet been reported.

Although much progress has already been made, for practical applications that require the manipulation of ultrasonic waves in the few megahertz range, an intrinsically low-loss solid matrix material is much desired. Aluminum (Al) is an excellent choice for matrix material, as it has no intrinsic loss in the ultrasonic frequency range. In this paper, we report on experimental demonstration of a complete band gap of 3D phononic crystals composed of body-centered tetragonal (bct) arrays of tungsten carbide (WC) spheres in Al matrix. Defect states in the band gap were also observed when a cluster of silicon nitride spheres was introduced into a 3D PC. Transmission wave field pattern further confirmed that the waves at the defect states frequencies were emitted from the defect cluster, which had a significantly larger diverging angle than the incident ultrasonic beam.

## 2. Experiment

The phononic crystals in the study consist of 1 mm diameter spherical WC scatterers imbedded in Al matrix. The bct crystal structure of our PC samples is shown in the insert of figure 2. The conventional lattice constants are 1.22 mm, 1.22 mm, and 1.0 mm, respectively. Commercial Al sheets with 99.5 % purity were used. The WC spheres are tungsten carbide cermet with cobalt as the binder (approximately 6 % in weight). The two materials have relatively large acoustic impedance mismatch: mass density $\rho = 2.7 \text{ g/cm}^3$, Young's Modulus $E = 70.3 \text{ Gpa}$ and Poisson Ratio $\nu = 0.33$ for Al; mass density $\rho = 14.0 \text{ g/cm}^3$, Young's Modulus $E = 587.9 \text{ Gpa}$ and Poisson Ratio $\nu = 0.22$ for WC. The defects are spheres of Silicon Nitride ($Si_3N_4$) in β structure with the same diameter. Its properties are in between Al and WC, namely mass density $\rho = 3.28 \text{ g/cm}^3$, Young's Modulus $E = 305.8$ GPa, and Poisson Ratio $\nu = 0.286$. We select Al as the matrix material because Al has much lower ultrasonic absorption in the frequency range of interest (0.5 MHz ~ 5 MHz) as compared to polymers such as epoxy [17], and Al has low acoustic impedance and low melting temperature as compared to other metals such as copper, steel. The samples were first assembled by hand in the ΓX direction in a mold. One layer in the samples consists of 22 rows of WC spheres with alternative assembly of 19 and 18 spheres in each row, with a nominal cross section area of $19.0 \times 19.2 \text{ mm}^2$ and layer interval of 0.866 mm (the nominal thickness of an n-layer PC could be calculated by $0.866 \times (n-1) + 1$ in mm). The volume fraction of the scatterers is ~70 %, which is within the desired range to form a full gap as predicted by theory [6]. A stack of aluminum sheets was placed on top of the empty PC below a plug. After the whole kit was heated up at 830 ℃ for 20 minutes in a vacuum oven, it was quickly placed under a laboratory press to force the molten Al into the space among the WC spheres before it solidified. This process can be repeated until the sample exhibits a transmission spectrum showing a well-defined band gap.

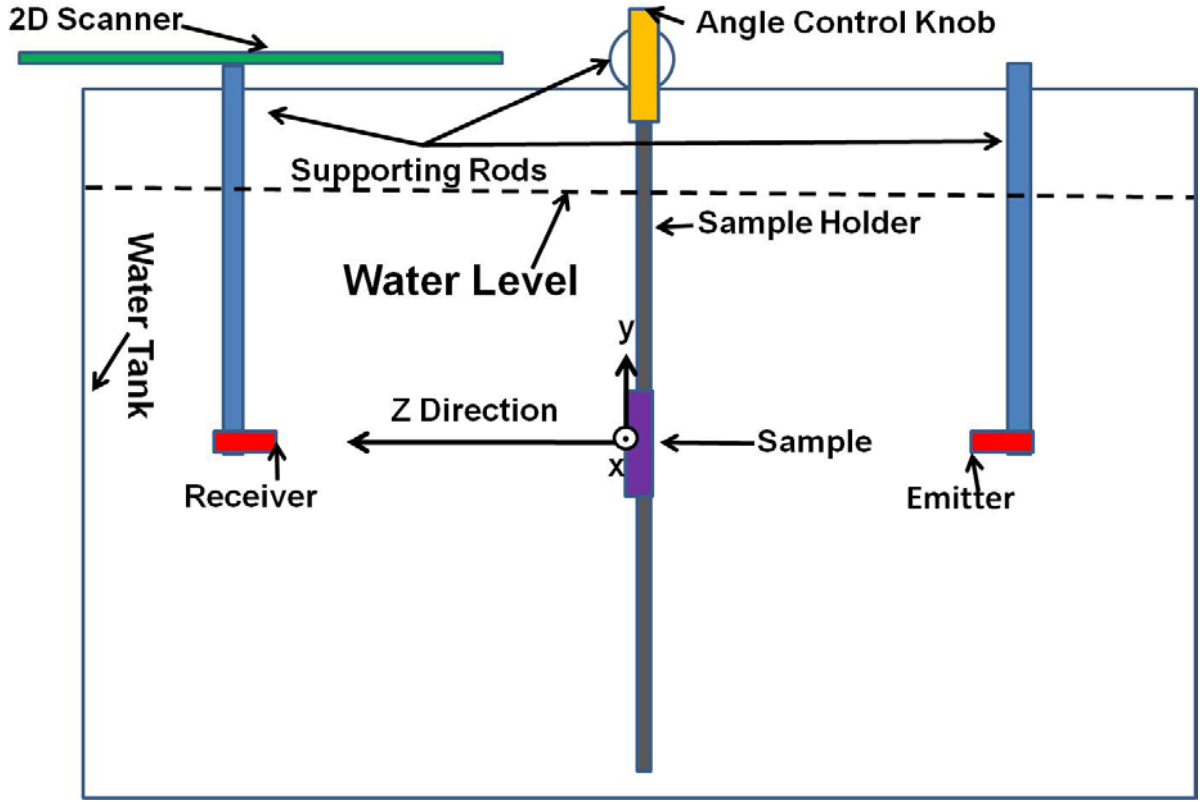

**Figure 1.** Schematic illustration of experimental measurement setup.

The transmission measurements using pulsed ultrasound technique were carried out in a water tank (size: $60 \text{cm} \times 40 \text{cm} \times 34 \text{cm}$) filled with distilled water as shown in figure 1. Anti-reflection pads were mounted on the entire walls of the tank. The sample was mounted on a sample holder between two piezoelectric transducers, one serving as the source and the other as the receiver. The sample holder had an opening for the PC mounting and prevented flank transmissions. The holder was mounted to a controllable knob, which could rotate around y-axis. Thus transmission measurements with different incidence angle could be made. The electric signal from the receiver was first recorded by a digital oscilloscope in the time domain and selected within a time window to keep only the signal transmitted directly through the sample while removing sporadic reflection signals arrived later. A transmission spectrum was obtained by measuring the waveform transmitted through the sample and the reference waveform measured without the sample. Fast Fourier Transforms (FFT) was then applied to the two waveforms, and the transmission amplitude was calculated as the ratio of the amplitude of the transmitted waveform over the reference waveform in the frequency domain, while the phase spectrum was obtained directly after shifting the transmitted waveform in the time domain by the amount for sound to travel in water over the distance outside the sample. Since the receiver was mounted to a 2D scanner, it could collect transmitted waveforms at various positions in the x-z plane, and after applying FFT, the spatial distribution of transmitted pressure field in the x-z plane could be mapped out at different frequencies.

### 3. Full band gap

The transmission spectra through a 9-layer PC are shown in figure 2. The three spectra correspond to the transmission spectra at normal incidence to the three pairs of surfaces of the PC along the [110], [001], and $[\bar{1}10]$ crystal directions. The thickness of the sample in the three directions are 7.72 mm ([110]), 19.35 mm ([001]), and 19.56 mm ($[\bar{1}10]$), respectively. All the spectra show a clear band gap which extends from 2.2 MHz to 3.3 MHz. The relative width of the band gap (the band width divided by the band gap central frequency) is around 40%, which is significantly larger than the one reported in Ref. 11 but smaller than that in Ref. 17. The impedance mismatch in our PC was smaller than that in Ref. 11, but yet it exhibited a much wider gap. This could be explained by the interaction of the transverse and longitudinal modes inside the PC [17]. The transmission contrast between the lower frequency passband and the band gap reaches 1000:1, or 60 dB, when using the transmission value around 0.5 MHz as reference. The contrast is 100:1, or 40 dB, when using the transmission value of the second passband around 4 MHz. The transmission minimum did not decrease as the PC layers were doubled, because the noise and leakage background of our system had been reached. We also measured transmission spectra with oblique incidence in both the ΓXZ and the ΓXX′ planes with incidence angle up to 30° with respect to the ΓX direction. All these spectra showed a common gap overlap in the 2.5 ~ 3.2 MHz region. Since all the measurements were taken in water and only longitudinal modes could propagate outside the PC, the measured gap would be wider than the real one which includes the transverse modes. However, there existed mode conversion inside the PC and at the PC-water interface, especially at oblique incidence. That is part of the reason why the oblique incidence measurements yielded a narrower gap. Combining all the transmission spectra obtained, we conclude that a 3D full gap exists in the WC/Al phononic crystals, at least for the longitudinal modes.

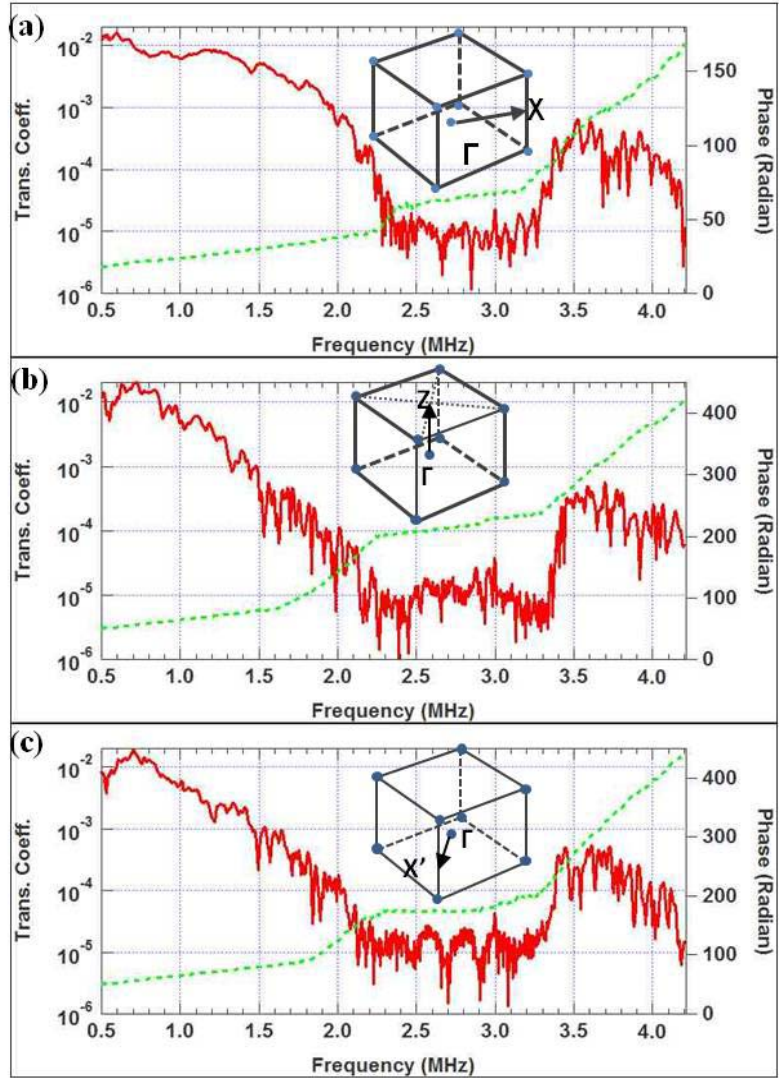

**Figure 2.** The transmission spectra of a 9-layer undoped phononic crystal in (a) the $[\bar{1}10]$ crystal direction, (b) the [001] direction, and (c) the [110] direction. The solid curves are the transmission amplitude and the dashed curves are the phase of the transmission. The insert depicts the crystal structure.

The phase spectra of the transmission (the dashed curves in figure 2) provide additional information about the wave propagation in the PC. Three types of regions in the frequency domain can be identified. The first type of regions are in the passbands where the dispersion is linear, i. e., the phase spectra are a straight line, such as in the frequency range below 1.5 MHz and between 3.6 MHz and 4.0 MHz in figure 2(b). The group velocity can be extracted from the phase spectra by using $v_g = d\omega/dk = L * d\omega/d\phi$, where $\omega$ is the angular frequency, $k$ the wave vector, $L$ the thickness of the PC and $\phi$ the phase. Note that L is different in different directions as described in the experiment section. The group velocities in the first passband are 4.01 km/s in the [001] direction, 4.34 km/s in the $[\bar{1}10]$ direction, and 4.22 km/s in the [110] direction. They are significantly lower than the sound speed in Al (6.42 km/s). In the second passband the corresponding group velocities are 0.605 km/s in the [001] direction, 0.515 km/s in the $[\bar{1}10]$ direction, and 0.546 km/s in the [110] direction. They are much lower than that in the first passband. It is also noted that the group velocities along the [110] and the $[\bar{1}10]$ directions are nearly equal, which is expected as the two crystal directions are equivalent, and are larger than in the [001] direction in the lower passband, but smaller in the higher passband. The second type of regions includes the top region of the first passband and the bottom region of the second passband next to the bandgap, where the phase spectra are curved in correspondence with the band bending near the bandgap. The third region is the bandgap, where the phase spectra are nearly flat, which indicates wave tunneling through the bandgap. The flat band region in the phase spectra matches well the significant drop in the transmission amplitude, confirming once more the existence of the bandgap. Theoretical modeling of the full band structure is under way. Comprehensive comparison with experimental results including the oblique incidence cases is beyond the scope of this paper and will be reported later.

## 4. Defect states

We next doped the PC by isolated $Si_3N_4$ spheres in 5-layer PCs, but could not observe any spectra that indicated the presence of localized defect states. This is probably because the scattering strength of a single $Si_3N_4$ sphere is not strong enough to cause wave localization due to the limited impedance contrast with WC. A planar cluster with one sphere in the middle of six surrounding ones was introduced in the 3$^{rd}$ layer of a 5-layer PC, as shown in the inset of figure 3, and the transmission spectrum in the ΓX direction showed clearly two transmission peaks inside the gap at 2.83 MHz and 3.07 MHz, as shown in figure 3 while there were no such peaks for the undoped 5-layer PCs. The presence of multiple peaks is due to the fact that the cluster has more than one vibration eigenmode. The full width at half maximum (FWHM) of the first peak is 0.04 MHz, corresponding to a quality factor of 71, while FWHM = 0.01 MHz for the second peak, corresponding to a quality factor of 307. The phase spectrum shows a phase change ~ π across the peaks at the two defects state frequencies, which indicates the resonant nature of the defects states. The transmitted pressure field of the impurity mode at frequency 3.07 MHz is shown in figure 4(a) together with that of a reference in 4(b) at the same frequency. The x and z directions are the same as described in figure 1. It is seen that near the sample the beam is significantly narrower than the reference beam directly from the source, and comparable to the size of the defect cluster. The spreading angle of the spatial FWHM of the transmitted field is 8.6°, which is significantly larger than the 5.0° spread of the reference beam. The field pattern at 2.83 MHz is similar and its spreading angle is 9.5°. Taking into account of the above observations, we conclude that localized defects states have been realized in 3D solid-in-solid PCs inside a full band gap.

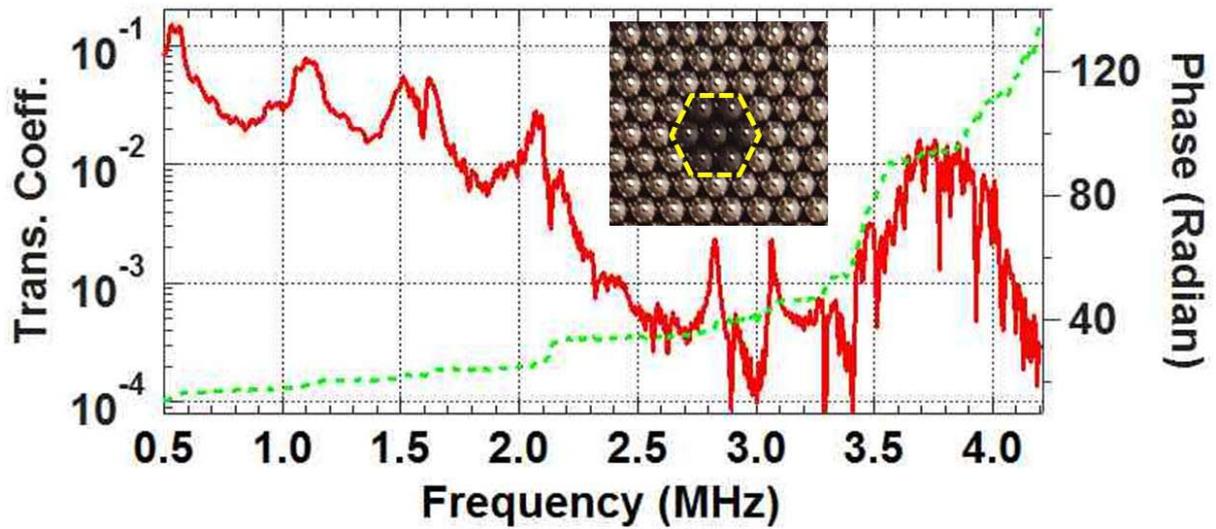

**Figure 3.** The transmission spectrum of a doped phononic crystal. The solid curve is the transmission amplitude and the dashed curve is the phase of the transmission. The insert is a picture of the defect cluster before it was covered by two layers of WC spheres.

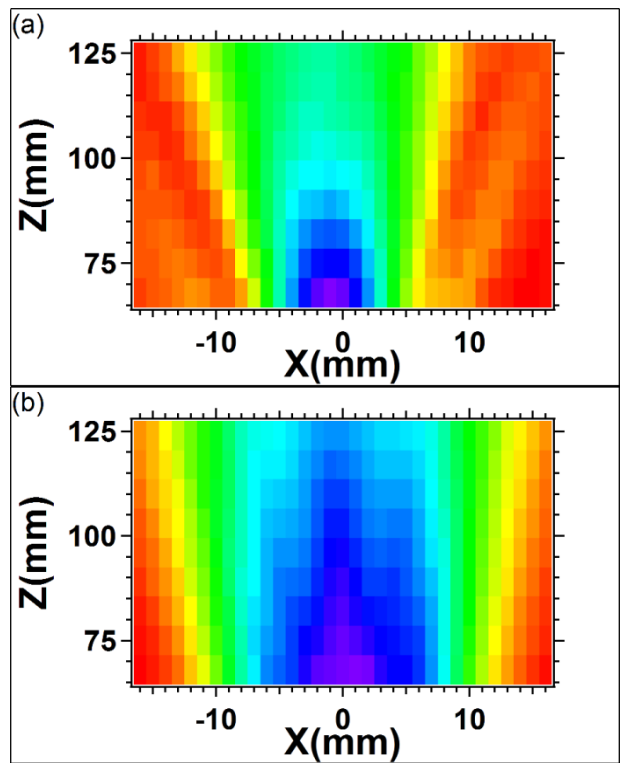

**Figure 4.** The pressure field pattern of (a) the transmitted waves at the defect state frequency of 3.07 MHz and (b) a reference directly from the source at the same frequency. The pressure amplitude decreases from color purple to red.

## 5. Conclusion

In this paper, 3D solid-in-solid PCs were experimentally studied. The non-absorption property of

Aluminum matrix makes this type of PC suitable for mega-hertz ultrasonic investigations. Full band gap of longitudinal modes with relative width 40% were demonstrated by underwater transmission spectra measurement. By introducing a cluster of $Si_3N_4$ defects, defect states were observed inside the gap frequency regime, which were also confirmed by the transmitted pressure field pattern mapping. Corresponding theoretical investigation is under way.

**Acknowledgement**

The authors thank P. Sheng for helpful discussions. The work was supported by GRF grant 606611 from the Hong Kong Government.